\documentclass[twocolumn,showpacs,preprintnumbers,amsmath,amssymb]{revtex4}

\usepackage{graphicx}
\usepackage{dcolumn}
\usepackage{bm}

\begin{document}


\title{Dynamic Phase Transitions in Superconductivity}

\author{Tian Ma}
\affiliation{Department of Mathematics, Sichuan University,
Chengdu, P. R. China%
}%

\author{Shouhong Wang}
 \homepage{http://www.indiana.edu/~fluid}
\affiliation{Department of Mathematics,
Indiana University, Bloomington, IN 47405
}%

\newcommand{\R}{\mathbb R}
\newcommand{\cC}{\mathcal C}
\newcommand{\C}{\mathbb C}
\newtheorem{theorem}{Theorem}
\newtheorem{acknowledgement}[theorem]{Acknowledgment}
\newtheorem{assumption}{Assumption}
\newtheorem{axiom}[theorem]{Axiom}
\newtheorem{claim}[theorem]{Claim}
\newtheorem{conclusion}[theorem]{Conclusion}
\newtheorem{condition}[theorem]{Condition}
\newtheorem{conjecture}[theorem]{Conjecture}
\newtheorem{corollary}[theorem]{Corollary}
\newtheorem{criterion}[theorem]{Criterion}
\newtheorem{definition}[theorem]{Definition}
\newtheorem{example}{Example}
\newtheorem{exercise}[theorem]{Exercise}
\newtheorem{lemma}[theorem]{Lemma}
\newtheorem{notation}[theorem]{Notation}
\newtheorem{problem}[theorem]{Problem}
\newtheorem{proposition}[theorem]{Proposition}
\newtheorem{remark}[theorem]{Remark}
\newtheorem{solution}[theorem]{Solution}
\newtheorem{summary}[theorem]{Summary}

\def\bt{\begin{thm}}
\def\et{\end{thm}}

\def\bl{\begin{lem}}
\def\el{\end{lem}}
\def\la{\label}

\def\bd{\begin{defi}}
\def\ed{\end{defi}}

\def\bc{\begin{cor}}
\def\ec{\end{cor}}

\def\bp{\begin{proof}}
\def\ep{\end{proof}}

\def\br{\begin{remark}}
\def\er{\end{remark}}

\newtheorem{thm}{Theorem}[section]
\newtheorem{lem}{Lemma}[section]
\newtheorem{defi}{Definition}[section]
\newtheorem{ex}{Example}[section]
\newtheorem{prop}[thm]{Proposition}
\newtheorem{xca}[thm]{Exercise}
\newtheorem{rem}{Remark}[section]
\newtheorem{cor}{Corollary}[section]
\newtheorem{method}[thm]{Method}

\date{\today}

\begin{abstract}
In this Letter, the dynamic phase transitions of the  time-dependent Ginzburg-Landau equations are analyzed  using a newly developed dynamic transition theory and a new classification scheme of dynamics phase transitions.  First, we demonstrate that there are two type of dynamic transitions, jump and continuous, dictated by the sign of a  nondimensional parameter $R$. This parameter is computable, and depends on the material property, the applied field, and  the geometry of domain that the sample occupies. Second, using the parameter  $R$,  precise analytical formulas for critical domain size, and for critical magnetic fields are derived. 
\end{abstract}

\pacs{74.20.-z, 74.20.De, 74.25.Dw}
\maketitle

One central problem in the theory of superconductivity is the
dynamical nature of the phase transition between a normal state,
characterized by a complex order parameter $\psi$ that vanishes
identically, and a superconducting state, characterized by the
order parameter that is not identically zero. In this Letter, we
address this problem by conducting rigorous theoretical analysis
on dynamic phase  transitions for the time dependent
Ginzburg-Landau (TDGL) model. 

{\bf TDGL model.} Let $\Omega\subset \R^n$ $(n=2,3)$ be a bounded domain,  $\psi :\Omega\rightarrow \C$  the order
parameter, $H:\Omega\rightarrow \R^n$ the magnetic field with $A$ being  the magnetic potential given by $H=\text{curl}\ A$,  and $H_a$ the applied field with potential $A_a$  such that  $\text{curl}A_a=H_a$. The  nondimensional TDGL equations  
for $(\psi, \mathcal A, \phi)$, with $\mathcal A$ being the deviation from the applied field, are given by:
\begin{equation}
\left.
\begin{aligned} 
&\frac{\partial\psi}{\partial t}+i\phi\psi
=-(i\mu\nabla +A_a)^2\psi +\alpha\psi
-2A_a\cdot{\mathcal{A}}\psi \\
& -2i\mu{\mathcal{A}}\cdot\nabla\psi -|{\mathcal{A}}|^2\psi
-\beta |\psi |^2\psi ,\\
&\zeta\left[\frac{\partial{\mathcal{A}}}{\partial
t}+\mu\nabla\phi\right]=-\text{curl}^2{\mathcal{A}}-\gamma A_a|\psi
|^2\\
& - \gamma {\mathcal{A}}|\psi |^2 
 -\frac{\gamma \mu}{2}i(\psi^*\nabla\psi -\psi\nabla\psi^*),\\
&\text{div}{\mathcal{A}}=0.
\end{aligned}
\right.\label{8.140}
\end{equation}
Here the nondimensional parameters are defined by 
\begin{align*}
&  \alpha =-2a\sqrt{b}m_sD/e^3_sh, &&  \beta =2m_sD/h,\\
&\mu =hD/\sqrt{b}e_s, &&  \zeta =4\pi\sigma le^2_s/c^2h, \\
& \gamma =4\pi e^2_s/m_sc^2l, &&\lambda =\lambda (T)=(m_sc^2b/4\pi e^2_s|a|)^{{1}/{2}},\\
& \kappa =\lambda /\xi, &&\eta =4\pi\sigma D/c^2,\\
\end{align*}
where $h$ is  the Planck
constant, $e_s$ and $m_s$ are the charge and mass of a Cooper pair,
$c$ is the speed of light, $\lambda =\lambda (T)$ is
the penetration depth, $\xi (T)$ is  the coherence length, 
$\tau$ for the relaxation time, and  $\kappa =\lambda /\xi$ is  the
Ginzburg-Landau parameter.

The TDGL equations are supplemented with an initial condition for $(\psi, \mathcal A)$,  a  free-slip boundary condition for $\mathcal A$, and 
either the Neumann or the Dirichlet or the Robin boundary conditions for $\psi$; see de Gennes \cite{degennes}.
Here We shall see in later discussions that the parameter $\alpha$
plays a key role in the phase transition of superconductivity,
which is given in terms of dimensional quantities by \cite{degennes}:
\begin{equation}\alpha =\alpha
(T)=\frac{2\sqrt{b}m_sDN_0}{e^3_sh}\cdot\frac{T_c-T}{T_c},\label{alpha}
\end{equation} 
where $N_0$ the density of states at the Femi level, and $T_c$  the critical temperature where incipient superconductivity
property can be observed.

{\bf Dynamic transition.} As mentioned earlier, the study of dynamic phase transition problem for the TDGL equations  is based on  a new dynamic transition theory by the authors \cite{b-book, chinese-book}. The starting point of the theory is to put the TDGL equations into the perspective of an infinite-dimensional dynamical systems as follows:
\begin{equation}
\frac{du}{dt} = L_\lambda u +G(u,\lambda), \qquad u(0) = u_0. \label{1}
\end{equation}
where $u: [0, \infty) \to H$ is the unknown function, $H$ is a Hilbert space, $L_\lambda$ is a linear operator, $G(u,\lambda)$ is a nonlinear operator, and $\lambda$ is the system parameter. Then under proper physical conditions, the dynamical transitions of a basic state of (\ref{1}) at a critical parameter $\lambda_0$, where the principle of exchange of stabilities hold true, can be classified into three categories: continuous as shown in Figure \ref{f8.22}, jump as shown in Figure \ref{f8.23}, and mixed; see the above references for further details. This theory has been applied to the TDGL 
equations (\ref{8.140}) to characterize the dynamic transition, will be be used in the analysis hereafter in this Letter.

{\bf Eigenvalue problem:} Consider the following equation
\begin{equation}
(i\mu\nabla +A_a)^2\psi =\alpha\psi\ \ \ \ \forall x\in\Omega
,\label{8.148}
\end{equation}
with  one of the boundary condition for $\psi$.
There are an infinite real eigenvalue sequence of
(\ref{8.148}) as
$0\le \alpha_1< \alpha_2 < \cdots < \alpha_k < \cdots \to +\infty.
$
The eigenvalues of (\ref{8.148}) always have even multiplicity.
In this Letter, we  consider only  the case where the first eigenvalue $\alpha_1$
of (\ref{8.148}) has multiplicity two, i.e., $\alpha_1$ is complex
simple eigenvalue. We let 
\begin{equation}
e=e_{1}=\psi_{11}+i\psi_{12},\ \ \ \
e_{2}=-\psi_{12}+i\psi_{11}.\label{8.152}
\end{equation}
be the eigenvectors corresponding to $\alpha_1$.

{\bf A necessary condition:} In superconductivity, the parameter $\alpha$ in (\ref{8.140}) can
not exceed a maximal value $\alpha (T)\leq\alpha (0)$ because
$T\geq 0$. Hence, a necessary condition for the possible phase transition from the normal state to  superconducting states is:
\begin{equation}
\alpha_1<\alpha (0)=\frac{2\sqrt{b}m_sDN_0}{e^3_sh},\label{8.158}
\end{equation}
where $\alpha_1$ is the first eigenvalue of (\ref{8.148}).

For the case where the Neumann boundary condition $\partial \psi/ \partial n=0$  on $\partial \Omega$,  the
first eigenvalue $\alpha_1=0$, which is
independent of $\Omega$, the geometry of  the sample. Therefore, the
condition (\ref{8.158}) always holds true.

However  for the Dirichlet and the Robin boundary conditions, 
the situations are  different. We know that the first eigenvalue $\alpha_1$  depends on $\Omega$. We have for example $\lim_{|\Omega |\to 0} \alpha_1=\infty.$
Hence, the condition (\ref{8.158})
implies that for the cases with either the Dirichlet or Robin boundary conditions, including the case where  the sample is enclosed by a
magnetic material or a normal metal, the volume of a sample must
be greater than some critical value $|\Omega |>V_c>0$. Otherwise
no superconducting state occurs at any temperature. 

Physical theory and experiments show that there is a critical
applied magnetic field $H_c$ by which a superconducting state will
be destroyed, and $H_c$ satisfies the following approximate
equation near the critical temperature $T_c$ which is given by
(\ref{alpha})
\begin{equation}
H_c(T)=H_0(1-T^2/T^2_c).\label{8.184}
\end{equation}

Equation (\ref{8.184}) is an empirical formula. A related equation
is the equation of critical parameter:
\begin{equation}
\alpha (T)=\alpha_1(H_a,\Omega ),\label{8.185}
\end{equation}
where $\alpha (T)$ is the parameter in (\ref{8.140}) and
$\alpha_1=\alpha_1(H_{\alpha},\Omega )$ the first eigenvalue of
(\ref{8.148}). It is expected that the applied
magnetic field $H_a$ satisfying (\ref{8.185}) is the critical
field $H_c$.

{\bf Definition of $R$:} We start with the introduction of a crucial
physical parameter, which completely determines the dynamical
properties of the phase transition of the Ginzburg-Landau
equations.

For the first eigenfunction $e$, the following Stokes problem has  a unique solution:\begin{equation}
\left.
\begin{aligned} 
& \text{curl}^2{\mathcal{A}}_0+\nabla\phi
=|e|^2A_a+\frac{\mu}{2}i(e^*\nabla e-e\nabla e^*),\\
& \text{div}{\mathcal{A}}_0=0,\\
& {\mathcal{A}}_0\cdot n|_{\partial\Omega}=0,\ \ \ \
\text{curl}{\mathcal{A}}_0\times n|_{\partial\Omega}=0.
\end{aligned}
\right.\label{8.159}
\end{equation}
Then we define a physical parameter $R$ as follows
\begin{equation}
R=-\frac{\beta}{\gamma}+\frac{2\int_{\Omega}|\text{curl}{\mathcal{A}}_0|^2dx}{\int_{\Omega}|e|^4dx}.\label{8.160}
\end{equation}

From (\ref{8.159}) and (\ref{8.160}) it is easy to see that the
parameter $R$ is independent of the choice of the first
eigenvectors of (\ref{8.148}) and $H_0=\text{curl}{\mathcal{A}}_0$
given by (\ref{8.159}) depend on the applied magnetic potential
$A_a$ and the geometric properties of $\Omega$, the parameter $R$
is essentially a function of $A_a, \Omega$ and physical
parameters $\beta ,\gamma,\mu$.

As mentioned earlier, there are two phase transitions:  Type-I and
Type-II, determined  by a simple parameter $R$ defined by (\ref{8.160}).
This parameter $R$ links the superconducting behavior with the
geometry of the material, the applied field and the physical
parameter's.

An equilibrium state $(\widetilde{\psi},\widetilde{{\mathcal{A}}})$
of the TDGL equations (\ref{8.140}) is called in the normal state
if $\widetilde{\psi}=0$, and it is called in the superconducting
state if $\widetilde{\psi}\neq 0$. A solution $(\psi
,{\mathcal{A}})$ of (\ref{8.140}) is said in the normal state if
$(\psi ,{\mathcal{A}})$ is in a domain of attraction of a normal
equilibrium state, otherwise $(\psi ,{\mathcal{A}})$ is said in the
superconducting state.

{\bf Dynamic phase transition for $R<0$:} 
 By the phase
transition theorems obtained by the authors \cite{mw05a}, the critical temperature $T^1_c$ of
superconducting transition satisfies $T^1_c<T_c$ where $T_c$ is as
in (\ref{alpha}) and $T=T^1_c$ satisfies (\ref{8.185}). It is
known that
$\lim_{|A_a|\to\infty} \alpha_1(H_a,\Omega =\infty, $ 
which  implies that the applied magnetic
field $H_a$ can not be very strong for superconductivity as
required by condition (\ref{8.158}).

We have shown in \cite{mw05a} that when $\alpha >\alpha_1$, the equations
(\ref{8.140}) bifurcate from $((\psi ,{\mathcal{A}}),\alpha
)=(0,\alpha_1)$ to a cycle of steady state solutions
$(\psi_{\alpha},{\mathcal{A}}_{\alpha})$ which is an attractor
attracting an open set $U \setminus \Gamma\subset H$. 
From the physical point of view,
this theorem leads to the following properties of superconducting
transitions in the case where $R<0$:

{\sc First}, when the control temperature decreases (resp.
increases) and crosses the critical temperature $T^1_c$, there
will be a phase transition of the sample from the normal to
superconducting states (resp. from the superconducting to normal
state). 

{\sc Second}, when the control temperature
$T>T^1_c$, under a fluctuation deviating the normal state, the
sample will be soon restored to the normal state. In addition,
when $T<T^1_c$, under a fluctuation deviating both the normal and
superconducting states, the sample will be soon restored to the
superconducting states. 

{\sc Third}, in general, the supercurrent
given by
$$J_s(\alpha )=-\gamma
(A_a+{\mathcal{A}}_{\alpha})|\psi_{\alpha}|^2-\frac{\gamma\mu}{2}i(\psi^*_{\alpha}\nabla\psi_{\alpha}-
\psi_{\alpha}\nabla\psi^*_{\alpha})$$ is nonzero, i.e.,
$J_s\not\equiv 0$ for $T<T^1_c$ $(\alpha_1<\alpha )$. 

{\sc Fourth}, the order parameter $\psi_{\alpha}$ and supercurrent
$J_s(\alpha )$ depend continuously on the control temperature $T$
(or the parameter $\alpha$):
$$\psi_{\alpha}\rightarrow 0,\ \ \ \ J_s(\alpha )\rightarrow 0\ \
\ \ \text{if}\ T\rightarrow (T^1_c)^-\ (\text{or}\
\alpha\rightarrow\alpha_1^+).$$ 

{\sc Fifth}, the superconducting
state of the system is dominated by the lowest energy
eigenfunction of (\ref{8.148}) in the  sense that 
$\psi =C |\alpha
-\alpha_1|^{{1}/{2}} e +o(|\alpha
-\alpha_1|^{{1}/{2}}).$

{\sc Sixth}, the phase transition is of second order
in the Ehrenfest sense with the critical exponent $\beta
={1}/{2}$, and its phase diagram is as shown in Figure \ref{f8.22}:
\begin{figure}[hbt]
  \centering
  \includegraphics[width=0.3\textwidth]{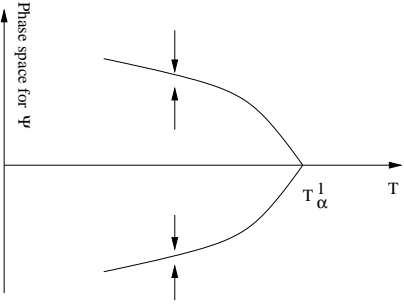}
  \caption{The continuous transition with $R <0$.}\la{f8.22}
 \end{figure}

{\bf Dynamic phase transitions with $R>0$:}
Consider a material described by
the TDGL model with $R>0$. There are two transition temperatures
$T^0_c$ and $T^1_c$  $(T^0_c>T^1_c)$ such that
$$\alpha (T^i_c)=\alpha_i\ \ \ \ (i=0,1)\ \ \ \ \text{with}\ \ \ \
\alpha_0<\alpha_1,$$ 
and the following phase transition properties
hold true:

{\sc First}, when the control temperature $T$ decreases and
crosses $T^1_c$ or equivalently $\alpha$ increases and crosses
$\alpha_1$, the stability of the normal state changes from stable
to unstable. 

{\sc Second}, when $T^1_c<T<T^0_c$ $(\alpha_0<\alpha
<\alpha_1)$, physically observable states consist of the normal
state and the superconducting states in $\Sigma^2_{\alpha}$ (see
Figure 8.20). When $T<T^1_c$ $(\alpha_1<\alpha )$, physically
observable states are in $\Sigma^2_{\alpha}$. 

{\sc Third}, when the control temperature $T$ is in the interval
$T^1_c<T<T^0_c$ (or $\alpha_0<\alpha <\alpha_1)$, the
superconducting states in $\Sigma^{\alpha}_1$ are unstable, i.e.,
with a fluctuation deviating a superconducting state in
$\Sigma^{\alpha}_1$, transition to either the normal state or a
super-conducting state in $\Sigma^{\alpha}_2$ will occur.

{\sc Fourth}, at the critical temperature $T^0_c$
 (resp. at $T^1_c)$ of the phase transitions, there is a jump from
the superconducting states to the normal state (resp. from the
normal state to superconducting states). 

{\sc Fifth}, the other
energy-level eigenfunctions possibly have a stronger influence for
the superconducting states. 

{\sc Sixth}, in the temperature
interval $T^1_c<T<T^0_c$, phase transitions occur and are
accompanied with the latent heat to appear.

Based on the conclusions (1)-(4) above, we can draw the phase
diagram in Figure \ref{f8.23}, where the critical temperature $T_c$  in the
interval $T^1_c<T_c<T^0_c$ is the transforming point.
\begin{figure}[hbt]
  \centering
  \includegraphics[width=0.3\textwidth]{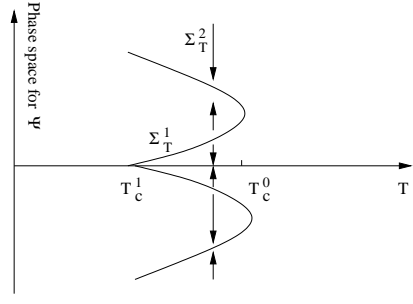} 
   \caption{The jump transition with $R > 0$.}\la{f8.23}
 \end{figure}

{\bf Critical Sample Size and Critical Magnetic Fields:}
Theories and experiments illustrate that the geometry of samples
$\Omega$ and applied magnetic fields $H_a$ have important
influences for the superconducting behaviors. In the following, we
shall apply the formula (\ref{8.160}) and eigenvalue equation
(\ref{8.148}) to discuss this problem.

For $0<L<\infty$  and $h > 0$, let 
$\Omega_0=D_0\times (0,h)\subset \R^3$, $x^{\prime}=(x_1,x_2)\in
D_0$, $0\in D_0$, and 
$$\Omega (L)=\{(Lx^{\prime},x_3)|\  x^{\prime} \in D_0, 0<x_3 <h\}.
$$ 
Let the applied field be given by 
$H_a=H_a(x^{\prime})=(0,0,H(x^{\prime}))$,  where $x^{\prime}=(x_1,x_2)\in
D_0.$ Then 
$H_a$  induces an applied  field
$\widetilde{H}_a$ on $\Omega (L)$ by
 $\widetilde{H}(y)=H(y/L)$ for any  $y=Lx^{\prime}$  with $x^{\prime}\in D_0.$
 
Let $H_a=\text{curl}A_a$, i.e., $H=\frac{\partial A_1}{\partial
x_2}-\frac{\partial A_2}{\partial x_1}$, where
\begin{equation}A_a=(A_1(x^{\prime}),A_2(x^{\prime}),0).\label{8.194}\end{equation}

Then we can show that  the nondimensional parameter $R=R(L,H)$   on $\Omega(L)$ is given by 
\begin{equation}
R(L,H)=-\kappa^2\mu^2+2L(p_3L^4+p_2L^2+p_1),\label{8.196}
\end{equation}
where  for any $0<L<\infty$, 
\begin{align*}
&
\left\{\begin{aligned}
&p_3=\frac{\int_{D_0}|\text{curl}A_0|^2dx^{\prime}}{\int_{D_0}|e|^4dx^{\prime}}>\delta
>0,\\
&\text{curl}^2A_0+\nabla\phi =|e|^2A_a\ \ \ \ \text{in }  x^{\prime}\in D_0,
\end{aligned}
\right.\\
&
\left\{\begin{aligned}
&p_1=\frac{4\mu^2\int_{D_0}|\text{curl}B_0|^2dx^{\prime}}{\int_{D_0}|e|^4dx^{\prime}},\\
&\text{curl}^2B_0+\nabla\phi =e_{12}\nabla e_{11}\ \
\ \ \text{in } x^{\prime}\in D_0,
\end{aligned}
\right.\\
& 
p_2=\frac{4\mu\int_{D_0}\text{curl}A_0\cdot\text{curl}B_0dx^{\prime}}{\int_{D_0}|e|^4dx^{\prime}}.
\end{align*}
Here $\delta >0$ is independent of $L$, and
$e(x^{\prime})=e_{11} + i e_{12}$ is the first eigenfuction of 
\begin{equation}
\left.
\begin{aligned} &(i\mu\nabla +L^2A_a)^2e=L^2\alpha e && \text{ in }  D_0,\\
&\frac{\partial e}{\partial n}=0    && \text{ on } \partial D_0.
\end{aligned}
\right.\label{8.197}
\end{equation}

Formula (\ref{8.196}) has many applications. In particular, we derive the  following effort of the domain and the applied field. It is easy to observe that 
$p_3$ is essentially a $|H|^2$ term,  $p_2$ is a $|H|$, and $p_1$  is a zeroth order of 
$|H|$.

First, by (\ref{8.196}), we derive that given a superconducting material, and an applied field $H_a\neq 0$, there is a critical scale $L_0>0$ such that the phase
transition in $\Omega (L)$ is a continuous  transition if $L<L_0$, 
and is a jump transition if 
$L>L_0$. In addition, $L_0$ is the unique real root of 
$$\alpha_3L^5_0+\alpha_2L^3_0+\alpha_1L_0=\frac{1}{2}\kappa^2\mu^2.$$
This theoretical conclusion is known experimentally  
for the  small sample $\Omega (L)$ (i.e., $1\gg L$) case.

Second, for Type I superconducting materials ($\kappa^2 < 1/2$), 
there are three critical magnetic fields $H_{c_1}<H_{c_2}<H_{c_3}$ such that
the following hold true:

\begin{enumerate}

\item If $0<H<H_{c_1}$, then the transition  is Type-I  and the superconducting state is in the Meissner state;

\item If $H_{c_1}<H<H_{c_2}$,  then transition is Type-II  and  in the Meissner state,

\item  If $\alpha_1(A_{c_2})=\alpha (0)$, then for any $H_{c_2}<H$,  it is in the normal state. 

\item If $\alpha_1(A_{c_2})< \alpha (0)$, then 

\begin{enumerate}

\item if $H_{c_2}<H<H_{c_3}$, then transition is Type-II  and  in the mixed state, and 

\item  if $H_{c_3}<H$, then it is in the normal state and $\alpha_1(A_{c_3})=\alpha (0)$.
\end{enumerate}

\end{enumerate}

Third, for type II superconductors,  the results are similar:

\begin{enumerate}

\item If $0<H<H_{c_1}$, then the transition is Type-I  and in the  Meissner state,

\item If $H_{c_1}<H<H_{c_2}$,  then transition is Type-I  and  in the mixed state,

\item  If $\alpha_1(A_{c_2})=\alpha (0)$, then for any $H_{c_2}<H$,  it is in the normal state.

\item If $\alpha_1(A_{c_2})< \alpha (0)$, then 

\begin{enumerate}

\item if $H_{c_2}<H<H_{c_3}$, then transition is Type-II  and  in the mixed state, and 

\item  if $H_{c_3}<H$, then it is in the normal state and $\alpha_1(A_{c_3})=\alpha (0)$.
\end{enumerate}

\end{enumerate}

We remark that the properties above are well
known by the Abrikosov theory and physical experiments. However,
here conclusions are derived rigorously using the recently developed dynamic transition theory, and the critical fields can 
be solved from (\ref{8.196}).


Fourth, for a small sample, there is a remarkable difference that the
first eigenvalue $\alpha_1(A_a)$ of  with
the Neumann boundary condition is bigger for small applied magnetic field
$H_a=\text{curl}A_a$. Hence the critical magnetic field
$H_c=\text{curl}A_c$ satisfying that $\alpha_1(A_c)=\alpha (0)$ is
very small, which implies that for a small sample, the
superconductivity is only in the Meissner state.

Finally we would like to mention that there have been extensive studies on bifurcation and stability 
analysis for superconductivity; 
see among others \cite{du1, BPT, sternberg, kaper, rubinstein}. 
The study in this article is based on a newly developed dynamic transition theory by the authors \cite{b-book, chinese-book}. With this new theory, many long standing problems in 
phase transition problems in science and engineering are becoming 
more accessible.
In particular, applications are made for various models 
from science and engineering, including, in particular, problems in statistical physics, 
classical and geophysical fluid dynamics.

\def\cprime{$'$}


\end{document}